\documentclass[procedia]{easychair}

\usepackage{doc}
\usepackage{makeidx}
\usepackage{diagbox}

\def\procediaConference{99th Conference on Topics of
  Superb Significance (COOL 2014)}

\title{Web Toolkit for Scientific Research: \\State of the Art and the Prospect for Development}

\titlerunning{Web Toolkit for Scientific Research \ldots}

\author{Stanislav P. Polyakov, Andrey P. Demichev and Alexander P. Kryukov\\
}

\institute{
  Skobeltsyn Institute of Nuclear Physics, Lomonosov Moscow State University, \\
  Leninskie gory, 1, Moscow, 119991, Russia\\
  \email{s.p.polyakov@gmail.com}, \email{demichev@theory.sinp.msu.ru}, \email{kryukov@theory.sinp.msu.ru}\\
 }

\authorrunning{Polyakov, Demichev and Kryukov}

\begin{document}

\maketitle

\keywords{web platforms, remote job submission, SaaS and PaaS models, remote software deployment, market of scientific application software}

\begin{abstract}
  The paper examines the current trends in designing of systems for convenient and secure remote job submission to various computer resources, including supercomputers, computer clusters, cloud resources, data storages and databases, and grid infrastructures by authorized users, as well as remote job monitoring and obtaining the results. Currently, high-perfomance computing and storage resources are capable of solving independently the majority of practical problems in the field of science and technology. Therefore, the focus in the development of a new generation of middleware shifts from the global grid systems to building convenient and efficient web platforms for remote access to individual computing resources. The paper examines the general principles of the construction and briefly describes some of the specific implementations of the web platforms.
\end{abstract}


%
%

\section{Introduction}
\label{sect:introduction}

The problem of creating the technology of remote job execution emerged within the conception of distributed computing \cite{1} in 1970s. In 1990s, as a result of the development of Internet and the widespread adoption of computer technology, distributed computing gained a boost in development, particularly within the grid paradigm \cite{2}, \cite{2a}. As a whole, the conception and the grid infrastructures created in the past years can be viewed as a comprehensive attempt to build distributed systems that would have to automatize the interaction of providers of data-processing services and their consumers. The most successful grid infrastructure is developing within ``The Worldwide LHC Computing Grid Project'' (WLCG, http://www.cern.ch/WLCG) in the European Organization for Nuclear Research (CERN, Geneva, Switzerland).

There are other types of geographically distributed systems, e.g. volunteer computing. In this case the client software periodically (typically during idle time when the local tasks are not performed) accesses the project server over the Internet, requesting the data for processing and sending the results back to the server. In this paper we will not discuss the web toolkits of this specific type (see e.g. survey \cite{3} and references therein).

With the growth of performance of individual resources (supercomputers, data storages, cloud systems \cite{4} etc.) the grid conception began to lose a significant part of its appeal. The fact is that the creation of a large-scale distributed computing grid infrastructure requires fairly high overheads both from the point of view of resource owner and administrators and from the point of view of grid users. In particular, the latter are required to fulfill rather complicated registration procedure, have to use the digital certificates and must posses rather high level of knowledge about the grid middleware. As a result, without a powerful unifying organizational structure, maintaining a cumbersome grid infrastructure proved to be very challenging, especially from the financial and administrative (overcoming the barriers between organizations owning the resources) points of view. For example, in the case of the WLCG project intended for support the one scientific mega device, namely, the Large Hadron Collider, such organizational structure is CERN. 

Currently, the high-performance compute and storage resources are able to solve most of the practical problems in science and technology separately. Therefore, the focus in the development of a new generation of middleware shifts to building convenient and efficient means for accessing individual computing resources. The principal development path of modern information technology aimed at facilitating access to resources for the end user and at reducing the time and financial costs is the use of web technologies. Within the software as a service (SaaS) model popular applications (e-mail, office applications, and so on) have been widely available for quite a long time. Currently the development of scientific web applications is booming around the world. In particular, today scientific services in the Internet provide the access to various data (for example, http://www.nature.com/sdata), scientific experimental facilities (for example, High Speed Networking with Subaru Telescope in Hawaii, http://www.naoj.org/), and educational materials (Virtual Learning Environment \cite{5}). 

In many cases researchers need to run a large number of similar computing tasks to solve some problem in a particular application area. Quite often, these tasks are performed by using already existing specialized application packages, the tasks being identical in form and differing only in the values of the input data (parameters). In this case, pre-installation and configuration of commonly used application packages on a computing resource and the availability of specialized web services as well as appropriate user web interfaces (SaaS model) allow owners of computing resources to increase efficiency of their use. The set of specialized web services and web application interfaces is called web platform for remote access to computing resources (other names used in the literature are: application-specific web portal, virtual laboratory, etc.). Each individual pre-arranged task (a launch of an application package, an access to a data storage, and so on) is called ``a tool''. With the access to the tools provided by a platform, a user only has to specify particular values of the input parameters or input files, and the rest of the task description is generated automatically. Thus the user can carry out the generating of the task, its submission, monitoring, and getting the results via a standard web browser. This approach proves to be particularly effective and convenient when the full research requires a set of application packages rather than just one of them. Moreover, often the output of some package is the basis for the generating of the task for the next application package in a series of stages of the research. Such series of tasks are called workflows. Besides the launch of computing packages, workflow may include other operations such as a query to relational database, visualization of the results or obtaining online data from an operating scientific facility. Carrying out of these operations can be made as simple as possible for end users by pre-configuring all the technical details (such as the Internet address of the resources, the specific formats of the query to a particular database or facility); in other words, creating a complete ``tool''. The user then only needs to formulate the essence of a specific request in a natural language.

General principles of constructing of the web platforms for remote access to computer resources are discussed in the next section. Section 3 briefly describes some of the specific implementations of web platforms.
\section{General Principles of Web Platforms}
\label{sect:general_principles}

The sets of features of web platforms can vary significantly. The following are the basic features of web platforms: (1) web platforms for job submission: remote submission, monitoring, and obtaining the job results; (2) web platforms for job submission and software installation: features from the item 1 plus remote installation and configuring of application packages, creating the complete tools for the use by other users; (3) web hubs: features from the items 1, 2 plus providing the features of professional social networks, for example, allowing to exchange the experience in the use of the platform tools, interaction with the developers, forming the rating of individual application packages/tools; (4) web platforms of application software market: features from the items 1, 2, 3 plus provision of information and computing web services for interaction between the providers and consumers of application packages for scientific research based on market principles (analogue of such application stores as AppStore, Google Play etc.).

Listed above are only the basic features of web platforms which categorize them by their purpose. In order to be fully operational, web platforms must include a variety of auxiliary services with appropriate features. The main functional requirements for the platform are divided into the following parts: management of user credentials granting the right to use the available resources; remote administration of the web platform via a web browser; job execution management; data files transfer management; tools (services) management. These requirements correspond to the web platforms with minimal functionality, namely the ones intended solely for remote job execution. In the case of web platforms with more general functionality  additional requirements arise. 

Generally the users of web platforms can operate in two modes: via a web browser and via a command line interface. Especially attractive for most users is the work via a web browser because in this case there is no need to install any additional software on their computers. Command line interface is aimed at advanced users and allows to automatize a number of actions.

The basis of the platforms are web service technologies and, in particular, the service-oriented architecture (SOA) \cite{6}, as well as the REST architectural style \cite{7}. Various authentication/authorization protocols are used to provide secure access to the resources via Internet with due regard to the user rights and service or resource policies. In addition to the well-known method based on the use of the login/password pair, protocols that can be used include OAuth (http://oauth.net), OpenID (http://www.openid.net), and those based on the public key infrastructure and X.509 standard certificates (http://www.ietf.org/rfc/rfc3820.txt). Authorization with the role-based determination of user access rights to the resources is usually carried out through the mechanism of dynamic mapping of users to the local resource accounts with the appropriate permissions.

If a web platform provides a service to install application software (including web hubs), one can use different types of resources provided that application software is deployed by the resources administrators. In this case, the web platform may contain tools to facilitate interaction between software developers and resources administrators. For example, these tools can include means for converting application software (subject to certain rules for the API) into ``software as a service'' (SaaS). As a basis for the creation of such web platforms open source middleware HUBZero (http://hubzero.org; see also, e.g. , \cite{9}) can be used. An example of use of the HUBZero is the web hub in nanotechnology NanoHUB (http://www.nanohub.org), developed in the USA.

However, if a web platform (including web hubs and web platforms of application software market) provides a service for remote installation of the application software by users themselves, the type of the resources is practically limited to the cloud systems. Indeed, an independent remote installation of application software on supercomputers by users is highly problematic both in terms of security and from a technical point of view: for a deployment of a software the user must deeply know the features of the architecture and software of the specific supercomputer. A natural solution to these problems is the use of virtual machines \cite{10} that provide both safety (isolation of installed software) and feasibility of using the operating system required for the application software to be installed. Thus, since cloud infrastructure provides tools for managing virtual machines, the web platforms allowing software installation must provide access to the cloud systems. In this case at the stage of the software installation the ``Platform as a Service'' (PaaS) model is used.

\section{Examples of Web Platform Implementations}
\label{sect:examples}

Development of web platforms for remote access to application software for research are widely underway around the world. Examples of such developments are shown in Table~\ref{tab:examples} including application areas for which the platforms were designed.
\begin{table}[htp]
	\begin{centering}
		\begin{tabular}{|c|l|l|}
		\hline
        \textbf{No.} &	\textbf{Web platform}   & \textbf{Application area} \\
	    \hline
	    1 & Web platform of educational-methodical software &\\
	      & package ``Multiscale modeling in nanotechnology'' \cite{11}&\\
	      &  (Photochemistry Center of the Russian Academy of & \\
	      & Sciences (RAS); http://www.nanomodel.ru) & nanotech\\
	    \hline
	    2 & ``Personal virtual computer'' system \cite{12} (South Ural & \\
	      &  State University; http://supercomputer.susu.ac.ru/pvc) & engineering\\
	    \hline
	    3 & UniHUB, the technological platform of the National & \\
	      & ``University Cluster'' program \cite{13} (Institute for &\\
	      & System Programming of RAS; https://unihub.ru) &  multipurpose\\
	    \hline
	    4 & Computing cloud platform of the Ural Branch of & \\
	      & the RAS \cite{14} (Institute of Mathematics and & \\
	      & Mechanics of the UrB RAS) & Matlab\\
	    \hline
	    5 & Web portal of the supercomputer management system \cite{15} & \\
	      & (V.M. Glushkov Institute of Cybernetics of NAS of & \\
	      &  Ukraine; http://melkon.com.ua/ru/cms) & multipurpose\\
	    \hline
	    6 & Everest web platform \cite{16} (Institute for Information & \\
	      & Transmission Problems of the RAS; & \\
	      & http://everest.distcomp.org) & multipurpose\\
	    \hline
	    7 & Multifunctional instrumental and technological platform&\\ 
	      & for cloud computing support CLAVIRE \cite{17} (Saint & \\
	      & Petersburg State University of Information Technologies, & \\
	      & Mechanics and Optics; http://clavire.ru) & $\sim$ 50 applications \\
	    \hline
	    8 & nanoHUB, web hub in nanotechnology \cite{18} (consortium of &\\
	      & the US universities; http://www.nanohub.org) & nanotech \\
	    \hline
	    9 & eQUEUE, web platform for the remote job submission & \\
	      & (AdvancedWebMO Clustering Technologies, Inc.; & \\
	      & http://www.advancedclustering.com) & multipurpose\\
	    \hline
	   10 & Nucleonica, scientific web portal (Institute for Transuranium & \\
	      & Elements; http://www.nucleonica.net) & nucler physics\\
	    \hline
	   11 & WebMO web platform (Hope College, Holland, USA;& \\
	      &  http://www.webmo.net) & molecular physics\\
	    \hline
	   12 & Yabi web platform \cite{19} (Centre for Comparative Genomics,& \\
	      & Murdoch, Australia; https://ccg.murdoch.edu.au/yabi)i & bioinformatics \\
	    \hline
	   13 & e-Science Central web platform \cite{20} (Newcastle & \\
	      & University, UK; http://www.esciencecentral.co.uk) & multipurpose\\
		\hline
		\end{tabular}
		\caption{Examples of existing web platforms with application areas for which they were designed.}
		\label{tab:examples}
	\end{centering}
\end{table}

Basically, all the web platforms have the typical three-layer architecture. The first layer is the frontend that provides the user web interface; the second layer is the platform engine and the administration module that is responsible for job management, tool configuration, audit trails and user management; the third layer is a resource manager that exposes data and compute resources to the preceding layer. Also all the platforms have more or less sophisticated security infrastructure. However, each of the platforms have peculiarities both in details of their implementation and in operational features. In particular, «Personal virtual computer», UniHUB, computing cloud platform of the UrB RAS, the Everest web platform and eQUEUE provide the basic capabilities of the web platforms for job submission (first item in the list in the beginning of the sect. 2). The WebMO platform belongs to the same type and improves the accessibility and usability of  computational chemistry packages. The web portal developed in the Glushkov Institute of Cybernetics of NAS of Ukraine provides, besides the user-friendly job submission, a convenient service including the console mode for managing a supercomputer or cluster. The peculiarity of the platform developed in the Photochemistry Center of RAS is that it provides a set of facilities for multiscale modeling typically occuring in nanotechnology.
The very successful and developed nanoHUB project lays emphasis on the educational goals in the area of nanosciences. Similarly the portal Nucleonica has grown to become the leading online resource in the nuclear sciences and is particularly suitable for education and training of young scientists, engineers and technicians in the nuclear domain. 
Thus though the underlying basic architecture and the web technologies are rather typical for the majority of web platforms for remote access to computing resources, they are flexible enough to adapt the middleware for specific purposes or application areas.
Below as specific examples of the different types we briefly introduce the Russian project CLAVIRE as well as the Yabi and e-Science Central projects. 

\subsection{CLAVIRE Project}

The multifunctional instrumental and technological platform CLAVIRE is designed for the efficient management of computing, information and software resources of distributed heterogeneous computer infrastructures within the cloud computing model \cite{17}. It can be used to provide users via Internet with high-performance domain-specific services within the cloud computing model for the various needs of science, industry, business, and the social sphere with the possibility of utilization of existing distributed computing infrastructure resources (dedicated supercomputers, grid infrastructure, cloud media). Features of the application and configuring of the platform for specific application areas are determined in each case by the general needs of this area in widely available high-performance computing and the availability of application software capable of being provided in the form of cloud services. CLAVIRE basic application software toolkit includes over 60 software packages in such areas as hydroaerodynamics and structural resistance, nanotechnology and quantum chemistry, hydrometeorology, shipbuilding, analysis and modeling of social systems and transportation infrastructure, bioinformatics. Under the control of the CLAVIRE platform, workflows (composite applications) can be designed which permit to organize interaction between different resources presented as the services within computational environments.

Organization of the process of creation and execution of workflows under control of the platform CLAVIRE are carried out within the concept iPSE \cite{21} and comes down to the step-by-step formalization of sets of job descriptions in terms of the workflows. The upper level of the application description is the metaflow. Its individual blocks contain only descriptions of computing tasks required by the user in the form of implicit instructions. Thus, the compiled metaflow is a formal description of user task in terms of subject area without any indication of the conditions for its implementation. In addition to describing the actions and data necessary for the computing, the user can specify the criteria for selection of specific services, resources, data (such as execution time or high reliability), as well as additional restrictions and parameters of certain actions (e.g. the required accuracy of the result).

With this method of setting the initial data, designing the workflows is a process of gradual specification of metaflow up to creating the specific scripts for starting services in the cloud and their further execution. At the first stage of the design process of the composite application the user can, for example, choose classes of services that are available in the cloud. These classes of services are used at the next stage for the selection of the specific services. Next, a workflow with the preselected specific implementations of computing services is created, followed by the scheduling and the creation of the execution script.
CLAVIRE platform is developed in the Saint Petersburg State University of Information Technologies, Mechanics and Optics in 2010-2012.

\subsection{Yabi Project}

Yabi is a web platform providing transparent access to heterogeneous high-performance compute and storage resources \cite{19}. It provides the execution of workflows consisting of successive tasks such as accesses to databases, use of the results of these queries as input for computing tasks etc. Yabi has well developed and convenient administrative interface for configuring the ``tools'' (software and databases preinstalled on computing resources) and for controlling the user access to these tools. As a user interface to access Yabi both web browser and command line can be used. Command line interface can be used by experienced users for further automation of their computing tasks. Initially, Yabi platform was developed for biomedical applications, but it can be customized for various (preinstalled) application software for remote job submission to heterogeneous resources (supercomputers, clusters, cloud systems, grid).

When a user composes (constructs from individual tasks) or reuses a workflow and executes it, the Yabi frontend submits a high level description of the entire workflow to the workflow engine that controls its step-by-step execution. Yabi web interface (which is available via a web browser) provides authentication for registered users and presents three views: the design view which allows construction of new workflows, the jobs view which shows previously submitted workflows along with their parameters and input files, and the files view that represents the data storages accessible by the user.

Yabi administration module provides web interface that allows an authorized user (Yabi administrator) to manage all aspects of the web platform operation including the creation of new tools from application software preinstalled on the resources and controlling user access to various tools. Yabi resource manager provides the two types of services: compute services and data services, both types having plug-in architecture that allows them to communicate with various compute resources and file storage systems. 

Overall, Yabi web platform is a simple for end users and customizable for various (preinstalled) application software system for remote job submission to heterogeneous high-performance computing resources. Yabi is developed in the Centre for Comparative Genomics; Murdoch, Australia.

\subsection{e-Science Central Project}

e-Science Central (e-SC) web platform \cite{20} provides software as a service (SaaS) and platform as a service (PaaS), allows users to develop and upload new services, and supports the creation of professional social networks. The services are provided on the basis of the cloud computing technology. Using only a web browser, users can upload their data, share it controllably with their colleagues, as well as analyze them by means of the preinstalled software or their original software that can be uploaded to the computing resources and made available to other users. The web platform also allows to design and execute workflows composed of the sets of tasks. 

For intellectual property protection, version tracking is carried out: when a file is stored, if a previous version exists, then a new version is automatically created. This is important for allowing users to work with old versions of data, services, and workflows, ensuring the reproduction of experiments and allowing investigations into the effects of changes to data and analysis processes over time. e-SC provides fine-grained security control for data, services, and workflows based on both user groups and user-to-user connections within a social network.

e-SC has an in-browser workflow editor that allows users to build a workflow by dragging services (either uploaded by themselves or shared by other users) from the structured list and connecting them together. Workflow services contain input and output ports, which can then be linked together. The input and output ports are able to restrict the types of data that can be sent to them, meaning that only compatible ports can be connected. The user can then execute the workflow. The results are displayed within a web browser window and also stored within the e-SC file system for later use.

Thus, e-SC platform is intended both for researchers in various fields, making it easier to store, share, and analyze their data, and for developers to create new scientific services and applications. Important for these goals is the use of cloud computing technology with SaaS and PaaS models. e-SC can be deployed on both public clouds (e.g. Amazon EC2 and Microsoft Windows Azure) and private clouds. e-Science Central is developed in Newcastle University, UK (http://www.esciencecentral.co.uk).

\section{Conclusion}
\label{sect:conclusion}

In this relatively brief survey we could not present all of the existing and perspective developments. However, the presented examples are sufficient to show that the web platforms for remote access significantly increase the efficiency of distributed computing and storage systems as well as stand-alone high-performance computer resources by virtue of simplification and unification of the access to software resources while maintaining high degree of security of the system.

It is worth noting that the success in creating and using the web platforms, presented in this paper, does not deny the possibility of using for scientific research of other distributed computing technologies. In particular, as noted in the Introduction, in the case of large-scale experiments in high-energy physics (such as the experiments at the Large Hadron Collider), large-scale grid infrastructures may be the most appropriate solutions. Discussion of these issues is beyond the scope of this paper and we refer the interested reader to the extensive existing literature, e.g., \cite{22}. The same applies to other megaprojects, such as, European x-ray free electron laser (XFEL, http://xfel.eu) or the International Thermonuclear Experimental Reactor (ITER, http://www.iter.org).

A short summary for web platforms, discussed in this paper, is presented in Table~\ref{tab:comparison} where the presence (+) or absence (-) of some functional properties of the web platforms are shown.

\begin{table}[htp]
	\begin{centering}
		\begin{tabular}{|l|c|c|c|c|c|c|c|c|c|c|c|c|c|}
		\hline
		\diagbox{\textbf{\small{Functional}}\\\textbf{\small{properties}}}{\textbf{\small{Web}}\\\textbf{\small{platform}}}                                                                        & 1 & 2 & 3 & 4 & 5 & 6 & 7 & 8 & 9 & 10& 11& 12& 13 \\
		\hline
	\small{Web interface}                                              & + & + & + & + & + & + & + & + & + & + & + & + & + \\
		\hline
         \small{CLI}                                                  & - & + & + & + & + & + & + & + & + & + & + & + & + \\
 	\hline
   \small{Automated creation}                                       &   &   &   &   &   &   &   &   &   &   &   &   &  \\
   \small{of problem-oriented}                                    &   &   &   &   &   &   &   &   &   &   &   &   &  \\
   \small{interfaces}                                              & - & - & + & - & - & - & + & + & - & - & - & + & - \\         
     	 \hline
   	 \small{Automatic decision-}                                 &   &   &   &   &   &   &   &   &   &   &   &   &  \\
   	 \small{making system}                                           &   &   &   &   &   &   &   &   &   &   &   &   &  \\
   	 \small{(intellectual instructor}                                      &   &   &   &   &   &   &   &   &   &   &   &   &  \\
   	 \small{for a choice of a suitable}                                   &   &   &   &   &   &   &   &   &   &   &   &   &  \\
   	 \small{ software)}                                   & - & - & - & - & - & - & + & - & - & - & - & - & - \\    
     		\hline
   	\small{Visualization}	                                           & + & - & + & - & - & - & + & + & - & + & + & - & - \\
   		\hline
   	 \small{Monitoring of}                                            &   &   &   &   &   &   &   &   &   &   &   &   &  \\
   	 \small{a task progress}	                                      & - & - & + & -/+ & + & - & + & + & - & + & + & - & - \\
         	\hline
   	 \small{Interaction with}                                          &   &   &   &   &   &   &   &   &   &   &   &   &  \\
   	 \small{external storage}                                          & + & + & + & + & + & + & + & + & + & + & + & + & + \\
   	     	\hline
   	 \small{Support of cloud systems}                                  & - & + & + & + & - & + & + & + & + & + & + & + & + \\
        	 \hline
   	 \small{Support of grid systems}                                   & - & - & + & - & - & + & + & + & - & - & - & + & - \\    
   	    	 \hline
   	 \small{Submitting and}                                             &   &   &   &   &   &   &   &   &   &   &   &   &  \\
   	 \small{compillation of}                                            &   &   &   &   &   &   &   &   &   &   &   &   &  \\
   	 \small{a custom software}                                           & - & + & + & - & + & - & + & + & + & - & - & - & - \\    
    	 \hline
   	 \small{Workflow support}                                          & -/+ & - & - & - & - & + & + & - & - & -/+ & -/+ & + & + \\
	\hline
     \small{Multiscale modeling at}                                    &   &   &   &   &   &   &   &   &   &   &   &   &   \\         
     \small{the level of}                                            &   &   &   &   &   &   &   &   &   &   &   &   &   \\
      \small{the middleware}                               & + & - & - & - & - & - & - & - & - & - & - & - & - \\          
	      \hline
	 \small{Remote deployment of}                                      &   &   &   &   &   &   &   &   &   &   &   &  & \\
	 \small{a user software}                                           & - & - & + & + & + & + & - & + & - & - & - & - & + \\    
	      \hline	
		\end{tabular}
		\caption{Comparison of functional properties of the web platforms listed in Table~\ref{tab:examples}. Here the column numbers match the row numbers in the Table~\ref{tab:examples}. The sign -/+ means restricted support of the property.}
		\label{tab:comparison}
	\end{centering}
\end{table}

Table~2 shows that the multifunctional instrumental and technological platform for cloud computing support CLAVIRE and the web hub in nanotechnology nanoHUB have the most developed set of functional properties. However, it should be noted that while the CLAVIRE platform  focuses on scientific research, nanoHUB mainly serves educational purposes.

Further line of development of the web toolkit may be related not only with the quantitative increase in the number of web-based platforms for remote access and the expansion of scientific, engineering, and manufacturing areas in which they are used, but also with the improvement of the technology of remote deployment of new application software on resources interacting with the web platforms.

This approach will help to overcome an important problem associated with the use of the SaaS model in scientific areas, namely, limited set of application packages offered by SaaS providers. Often, these providers focus on mass servicing of single-type customers and scientific activity is beyond the scope of their interests. Currently, the provision of services for providers of application software in the context of scientific-oriented web platforms is not developed enough. Although some implementations (for example, e-Science Central) have services for remote application software deployment, they are still insufficient to ensure the creation of a web platform capable of performing the whole range of tasks characteristic for a free open market. The technology of creating such web platforms market of application software can be based both on the original solutions and on the synthesis and adaptation of the solutions used in research hubs (e.g., nanoHUB; nanohub.org), cloud and grid systems, as well as in on-line app stores. However, it seems that unlike the on-line app stores, the platform should not only provide information services for searching the tools needed by users, but also provide the feasibility of direct using of the necessary tools. Thus, the future web platforms will provide a single entry point both for web service providers and for their customers.

\subsection*{Acknowledgments}
\label{sect:acks}

This work is supported by the Ministry of Science and Education of Russian Federation, agreement No. 14.604.21.0146.

%
\label{sect:bib}
\bibliographystyle{plain}
\bibliography{WebTools_v4}

\begin{thebibliography}{10}

\bibitem{12}
K.V. Borodulin and A.P. Sviridov.
\newblock Implementation of ``personal virtual computer'' technology based on
  open source technology.
\newblock In {\em Proceedings of the International Scientific Conference on
  Parallel Computational Technologies. (PCT'2013)}, page 585, Chelyabinsk,
  2013. SUrSU Publishing Center.
\newblock In Russian.

\bibitem{21}
A.V. Boukhanovsky, S.V. Kovalchuk, and S.V. Maryin.
\newblock Intelligent software platform for complex system computer simulation:
  Conception, architecture and implementation.
\newblock {\em Izvestiya VUZov. Priborostroenie}, 52(10):5--24, 2009.
\newblock In Russian.

\bibitem{6}
T.~Erl.
\newblock {\em SOA Design Patterns}.
\newblock Prentice Hall, New York, 2009.

\bibitem{2}
I.~Foster and K.~Kesselman.
\newblock {\em The Grid, Blueprint for a New computing Infrastructure}.
\newblock Morgan Kaufmann Publishers, San Francisco, 1998.

\bibitem{2a}
I.~Foster and K.~Kesselman.
\newblock {\em The Grid 2, Blueprint for a New computing Infrastructure}.
\newblock Morgan Kaufmann Publishers, San Francisco, 2004.

\bibitem{22}
R.A. Gerber and H.J.~Wasserman (Eds.).
\newblock Large scale production computing and storage requirements for high
  energy physics: Target 2017.
\newblock Technical report, LBNL, University of California, 2012.
\newblock LBNL-6491E.

\bibitem{14}
M.L. Goldstein, A.V. Sozykin, and D.A. Ustalov.
\newblock Computing cloud platform of the urb ras.
\newblock In {\em Proceedings of the International Supercomputing Conference
  Scientific Service on the Internet: All Facets of Parallelism}, pages 79--81,
  Moscow, 2013. Publishing of the Moscow State University.
\newblock In Russian.

\bibitem{15}
A.L. Golovinski, A.L. Malenko, and L.F. Belous.
\newblock Web portal of supercomputer management system.
\newblock {\em Numerical Methods and Programming}, 11(7):130--136, 2010.
\newblock In Russian.

\bibitem{20}
H.~Hiden, S.~Woodman, P.~Watson, and J.~Cala.
\newblock Developing cloud applications using the e-science central platform.
\newblock {\em Phil Trans R. Soc.}, page 20120085, 2013.

\bibitem{19}
A.A. Hunter, A.B. Macgregor, T.O. Szabo, C.A. Wellington, and M.I. Bellgard.
\newblock Yabi: An online research environment for grid, high performance and
  cloud computing.
\newblock {\em Source Code for Biology and Medicine}, 7:1--22, 2012.

\bibitem{10}
E.S. James and N.~Ravi.
\newblock {\em Virtual Machines: Versatile Platforms For Systems And
  Processes}.
\newblock Morgan Kaufmann, San Francisco, 2005.

\bibitem{18}
G.~Klimeck, M.~McLennan, S.P. Brophy, G.B. Adams, and M.S. Lundstrom.
\newblock nanohub.org: Advancing education and research in nanotechnology.
\newblock {\em Computing in Science and Engineering}, 10(5):17--28, 2008.

\bibitem{17}
K.~V. Knyazkov, S.V. Kovalchuk, T.N. Tchurov, S.V. Maryin, and A.V.
  Boukhanovsky.
\newblock Clavire: e-science infrastructure for data-driven computing.
\newblock {\em Journal of Computational Science}, 3:504--510, 2012.

\bibitem{3}
E.~J. Korpela.
\newblock Seti@home, boinc, and volunteer distributed computing.
\newblock {\em Annual Review of Earth and Planetary Sciences}, 40:69--87, 2012.

\bibitem{9}
M.~McLennan and R.~Kennell.
\newblock Hubzero: A platform for dissemination and collaboration in
  computational science and engineering.
\newblock {\em Computing in Science and Engineering}, 12(2):48--52, 2010.

\bibitem{4}
P.~Mell and T.~Grance.
\newblock {\em The NIST Definition of Cloud Computing. Recommendations of the
  National Institute of Standards and Technology. Special Publication 800-145}.
\newblock NIST, Washington, 2011.

\bibitem{11}
V.~Mizgulin, S.~Goldstein, and R.~Kadushnikov.
\newblock ``cloud'' platform for nanotechnology research and development.
\newblock {\em Nanoindustry}, 5:60--64, 2011.
\newblock In Russian.

\bibitem{7}
L.~Richardson and S.~Ruby.
\newblock {\em RESTful Web Services}.
\newblock O'Reilly Media, Cambridge, 2007.

\bibitem{13}
O.I. Samovarov and S.S. Gaysaryan.
\newblock The web-laboratory architecture based on the cloud and the unihub
  implementation as an extension of the openstack platform.
\newblock {\em Proceedings of the Institute for System Programming},
  26(1):403--6420, 2014.
\newblock In Russian.

\bibitem{16}
O.~Sukhoroslov and A.~Afanasiev.
\newblock Everest: A cloud platform for computational web services.
\newblock In {\em Proceedings of the 4th International Conference on Cloud
  Computing and Services Science (CLOSER 2014)}, pages 411--416, Moscow, 2014.
  Science and and Technology Publications.

\bibitem{1}
A.S. Tanenbaum and M.~van Steen.
\newblock {\em Distributed Systems: Principles and Paradigms}.
\newblock Prentice Hall, Upper Saddle River, 2002.

\bibitem{5}
M.~Weller.
\newblock {\em Virtual learning environments: using, choosing and developing
  your VLE}.
\newblock Routledge, London, 2007.

\end{thebibliography}

\end{document}